# Gate-controlled transmission of quantum Hall edge states in bilayer graphene

Jing Li[1], Hua Wen[1], Kenji Watanabe[2], Takashi Taniguchi[2], Jun Zhu[1,3*]

[1]Department of Physics, The Pennsylvania State University, University Park, Pennsylvania 16802, USA.

[2]National Institute for Material Science, 1-1 Namiki, Tsukuba 305-0044, Japan.

[3]Center for 2-Dimensional and Layered Materials, The Pennsylvania State University, University Park, Pennsylvania 16802, USA.

*Correspondence to: jzhu@phys.psu.edu (J. Zhu)

**Abstract**

The edge states of the quantum Hall and fractional quantum Hall effect of a two-dimensional electron gas carry key information of the bulk excitations. Here we demonstrate gate-controlled transmission of edge states in bilayer graphene through a potential barrier with tunable height. The backscattering rate is continuously varied from 0 to close to 1, with fractional quantized values corresponding to the sequential complete backscattering of individual modes. Our experiments demonstrate the feasibility to controllably manipulate edge states in bilayer graphene, thus opening the door to more complex experiments.

The edge states of quantum Hall (QH) and fractional quantum Hall (FQH) effects are not only fascinating one-dimensional quantum fluid with rich dynamics of their own [1] but also provide access to the unconventional charge and statistics of the quasi-particle excitations of the bulk many-body ground states [2-4]. A well-known example is the even-denominator FQH state at filling factor $\nu = 5/2$ in GaAs quantum wells [5,6], where a ground state with non-Abelian excitations has long been hypothesized [7], yet experimental confirmation remains difficult and controversial [8-10]. The 5/2 state in GaAs is fragile and the electrostatic environment of high-quality GaAs samples is quite complex [11]. Recent technological advances have enabled remarkable strides in the quality of the two-dimensional electron gas (2DEG) in graphene [12,13]. Bilayer graphene, for example, exhibits a plethora of broken-symmetry QH, FQH and QH ferromagnetic states [14-19]. Importantly, even-denominator FQH states with large gaps of a few Kelvin have been observed [20-22]. The thin profile of a graphene device enables smaller and more precise nanostructures, such as demonstrated in our previous work on the quantum valley Hall kink states and valleytronic operations in bilayer graphene [23,24]. The simultaneous advances of sample quality and device fabrication techniques now enable more sophisticated edge state experiment in graphene. Past experiments have shown that naturally formed, smooth potential interfaces in a *p-n* [25-28] or *p-n-p* or *p-p'-p* junction [29-33] allow edge states to fully equilibrate. Spin polarization imposes a selection rule at low Landau levels [31,32]. Klimov et al observed partial equilibration at a *p-n* junction where a barrier is present although the barrier height is not tunable [34]. Recently, Zimmermann et al created a quantum point contact (QPC) geometry in graphene using a pair of top split gates and showed its control over the transmission of the edge modes [32]. This control, however, is less straightforward since carriers underneath the split gates cannot be depleted and produce edge states of their own that assist in tunneling. A clean QPC action, where a gate-tuned potential barrier controls the interaction between two quantum Hall edges, has not been realized in graphene.

In this Letter, we report on gate-controlled transmission of edge states between two lateral QH states in bilayer graphene. We use a dual split-gated structure to control the filling factor of the left and right QH states independently and a fifth gate to modulate the height of the tunnel barrier between the two. The tunneling resistance varies with the barrier height and exhibits plateaus that correspond to complete backscattering of individual edge states one by one. The experimental observations are quantitatively captured by finite element simulations of the device. This study is a proof-of-concept demonstration towards the construction of more sophisticated structures, such as a Fabre-Perot interferometer.

Figure 1(a) shows an optical micrograph of one of our devices (device 47) with its top and side view schematically shown in Figs. 1(b) and (c) respectively. The split junction is 70 nm in width and 300nm in length in both devices, 43 and 47. The devices are similar in structure and fabrication to those used to demonstrate the quantum valley Hall kink states [23]. In device 47, the gating efficiencies are respectively 8.04 and 6.00 × $10^{11}$ cm$^{-2}$V$^{-1}$ for the top (TL and TR) and bottom (BL and BR) split gates. They correspond to thicknesses of 20 nm and 28 nm respectively for the top and bottom hexagonal-Boron Nitride (h-BN) dielectric layers, with $\varepsilon = 3.0$ [23].

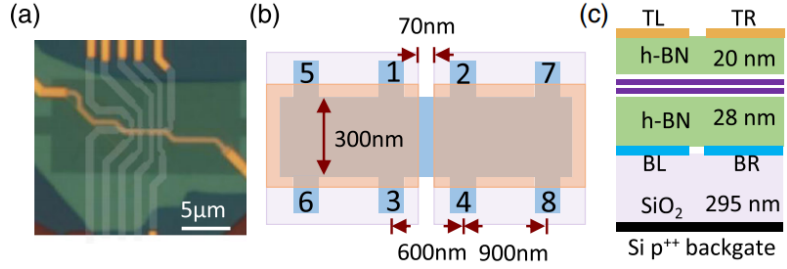

FIG. 1. (a) An optical micrograph of device 47. The bottom split gates are made of multi-layer graphene (dark squares). The top split gates are Au. The bilayer graphene is etched into a multi-probe Hall bar highlighted in white. The top and bottom h-BN sheets appear in green and dark blue shades respectively. (b), (c) Schematics of the top and side views of device 47. Orange and light purple shades illustrate the top and bottom split gates respectively. The parameters of device 47 are given in the diagrams.

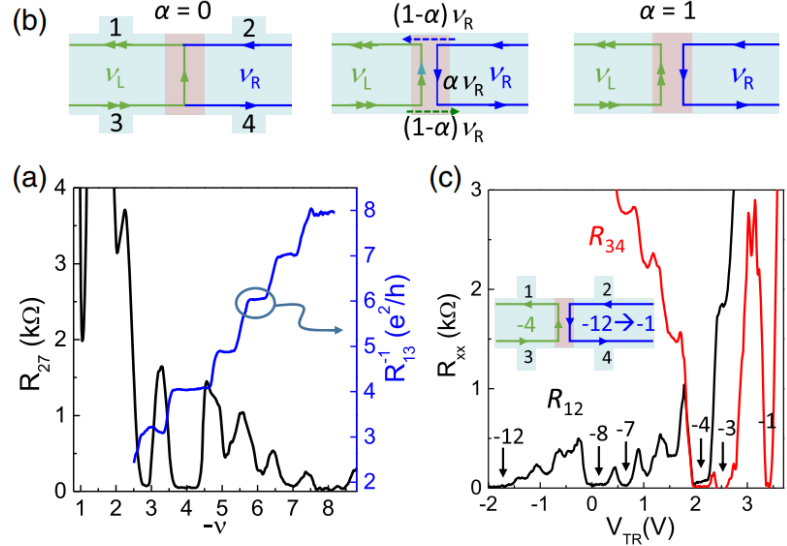

FIG. 2. (a) $R_{27}$ (black curve, left axis) and $R_{13}^{-1}$ (blue curve, right axis) vs filling factor $v$ measured on the right and left side of device 47 respectively. $R_{27}(v_R)$ is measured by sweeping $V_{TR}$ while fixing $V_{TL} = 0$ V. $R_{13}(v_L)$ is measured by sweeping $V_{TL}$ while fixing $V_{TR} = 0$ V. In both measurements, $V_{BL} = V_{BR} = -4.5$ V and $V_{Si} = -20$ V. From device 47. (b) The edge states flow diagram in a unipolar junction with $|v_L| > |v_R|$. The left, middle and right panels show perfect transmission, partial backscattering and complete backscattering of the right side edge states respectively. (c) $R_{12}$ (black trace) and $R_{34}$ (red trace) as a function of $V_{TR}$. $V_{BR} = -5.0$ V. $v_R$ varies from -12 to -1 as the inset shows. $V_{Si} = -30$ V. Arrows indicate the integer fillings of $v_R$. $v_L = -4$ is fixed by setting $V_{BL} = -4.02$ V, $V_{TL} = 1.4$V. From device 43.

Landau levels (LLs) form when a perpendicular magnetic field is applied [17,19]. Using the four split gates, we can vary the filling factors $v_L$ and $v_R$, and the displacement electric fields $D_L$ and $D_R$ of the left and right QH states independently. We pass a constant current through the entire device and measure $R_{xx}$ and $R_{xy}$ of each side, as well as $R_{xx}$ across the junction simultaneously using standard low-frequency lock-in techniques. Early onset of symmetry-broken integer QH states and the appearance of FQH states attest to the reasonably high quality of our devices. (See Supplementary Figure S1) Data presented here are acquired at $B = 18$ T and $T = 0.3$ K. Figures 2(a) plots examples of $R_{xy}^{-1}$ and $R_{xx}$ obtained on device 47. Both sides exhibit well-resolved integer QH states in the $p$-type carrier regime. Using these measurements, we select well-developed QH regimes for subsequent edge tunneling measurements. Results reported here focus on unipolar *p-p* junctions.

When both sides of the junction are positioned at integer filling factors as illustrated in Fig. 2(b), edge states propagate at the sample boundary and interact along the line junction, the potential of which is controlled by $V_{Si}$ applied to the doped silicon backgate in Fig. 1(c). Here we have set $|\nu_L| > |\nu_R|$. The junction may backscatter a fraction of the edge states from the right, as the middle panel of Fig. 2(b) shows. The backscattering rate $\alpha$ is controlled by the height of the junction potential. $\alpha = 0$ in the left panel corresponds to the situation of perfect transmission while $\alpha = 1$ in the right panel depicts the situation of complete backscattering.

Using the Landauer-Büttiker formula for edge state transport, we can relate $\alpha$ to $R_{xx}$ measured across the junction, e.g. $R_{12}$ or $R_{34}$ in Fig. 2(c) and obtain

$$R_{34} = \frac{\alpha}{1-\alpha} \frac{1}{\nu_R} \frac{h}{e^2}, \qquad (1)$$

which is along the bottom side of the sample where the edge states flow from left to right (assuming $|\nu_L| > |\nu_R|$). Similarly, along the top side of the sample, we find

$$R_{12} = \left[ \frac{1}{1-\alpha} \frac{1}{\nu_R} - \frac{1}{\nu_L} \right] \frac{h}{e^2}. \qquad (2)$$

The expressions of $R_{12}$ or $R_{34}$ exchange with one another in the case of $|\nu_L| < |\nu_R|$ or when the direction of the magnetic field is reversed.

Eqs. (1) and (2) enable us to measure the edge state backscattering rate $\alpha$ directly, similar to past studies in GaAs [35-37]. $R_{34}$ vanishes in the case of $\alpha = 0$, where a $|\nu_R|$ number of edge states flow through the junction along both top and bottom sides of the sample without backscattering. A finite $R_{34}$, together with simultaneously vanishing $R_{xx}$ of the bulk QH states, indicates backscattering at the junction.

An example of a transparent junction is given in Fig. 2(c) using data in device 43. Here, we set $\nu_L = -4$ and sweep gate voltage $V_{TR}$ to change $\nu_R$ from -12 to -1. $V_{Si}$ is fixed at -30 V. Both $R_{12}$ (black line) and $R_{34}$ (red line) vs $V_{TR}$ are plotted. From $-4 \leq \nu_R < -1$, Eq. (1) describes $R_{34}$ whereas from $-12 < \nu_R \leq 4$, Eq. (1) describes $R_{12}$ instead. We see immediately that $\alpha = 0$ when $\nu_R$ is at the integer fillings of -1, -3, -4, -7, -8 and -12, i.e. the junction is transparent. In fact, non-zero $R_{12}$ (or $R_{34}$) observed at other integer fillings of $\nu_R$ ($\nu_R = -2, -5, -6$ and so on) is likely due to contributions from non-zero $R_{xx}$ of the bulk, as $R_{27}$ shown in Fig. 2(b) suggests. As we will show in the simulations, $V_{Si} = -30$ V corresponds to a *p-p′* junction with a smooth interface potential profile, similar to what's studied in Refs. [25-28].

Next, we investigate the effect of the junction potential on $\alpha$. In Fig. 3 (a), we plot $R_{34}$ as a function of $V_{Si}$, while fixing both filling factors $\nu_L$ and $\nu_R$ to be $(\nu_L, \nu_R) = (-4, -2)$. Here $D_L = -0.2$ V/nm is fixed on the left while different traces correspond to different values of $D_R$. The D-field controls the energies of the bilayer graphene LL spectrum [19].

A large $D$-field promotes the splitting at $\nu = \pm 1$ and $\pm 3$. A quantitative LL diagram between $-4 < \nu < +4$ is given in Fig. S2 of the Supplementary Material. In Fig. 3(a), we label the four distinct regimes for the dark cyan trace corresponding to $D_R = -0.2$ V/nm. In Regime II, $R_{34}$ is close to zero, which indicates the perfect transmission of both edge modes of $\nu_R$ across the junction. As $V_{Si}$ increases further, $R_{34}$ becomes finite and eventually reaches large values (Regimes III and IV). This behavior corresponds to the increase of the potential barrier between the two QH states as $p$-type carriers in the junction are increasingly depleted, eventually causing all edge states to backscatter completely. To the left of Regime II, a moderate increase of $R_{34}$ is also observed when $V_{Si}$ becomes very negative and the junction becomes heavily $p$-doped. This is labeled as Regime I in the plot. Finite bulk conduction across the junction is likely responsible.

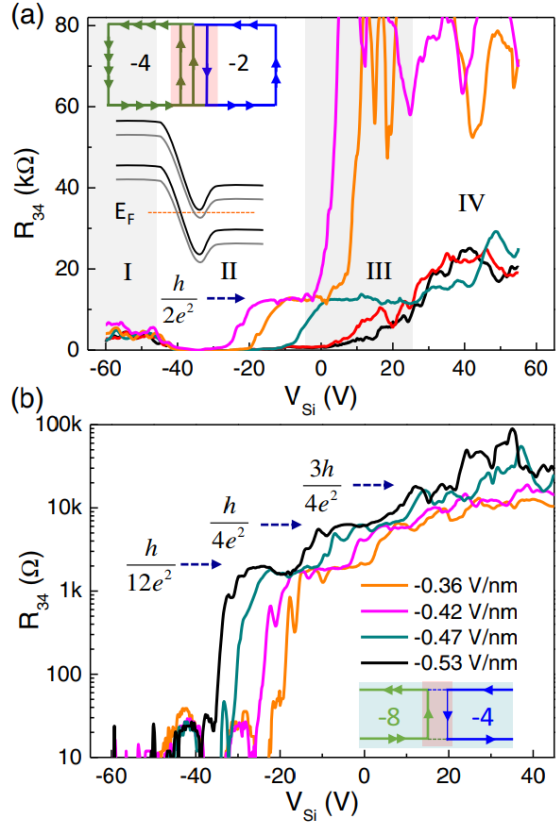

FIG. 3. (a) $R_{34}$ as a function of $V_{Si}$. From left (magenta) to right (black): $D_R = -0.35, -0.30, -0.20, -0.10$ and $-0.05$ V/nm. $D_L = -0.2$ V/nm for all traces. $(\nu_L, \nu_R) = (-4, -2)$. Positive $D$ corresponds to an electric field pointing up. See Ref. [23] for the definition of $D$ and how we control $\nu$ and $D$ independently. Transmission regimes I-IV are marked for the $D_R = -0.20$ V/nm trace (dark cyan). The dashed arrow indicates resistance value of $h/2e^2 = 12.9$ k$\Omega$. Insets illustrate the flow of the edge states and the potential profile across the junction in regime III. (b) $R_{34}$ vs $V_{Si}$ in the case of $(\nu_L, \nu_R) = (-8, -4)$ as the inset illustrates. $D_L = -0.21$ V/nm, $D_R$ varies as labeled in the plot. Dashed arrows mark resistance values of $h/12e^2$, $h/4e^2$ and $3h/4e^2$. From device 47.

What is most striking in Fig. 3(a) is the appearance of a wide plateau in $R_{34}$ (Regime III in the plot). The resistance value of the plateau is close to $h/2e^2$, which corresponds to $\alpha = 1/2$ in Eq. (1). The appearance of the quantization is intuitive when considering the evolution of the bulk LLs inside the junction. As the insets of Fig. 3(a) show, when $E_F$ resides between the first and second LLs inside the junction, one edge mode is completely backscattered while the other fully transmits through the junction. As our data in Fig. 3(a) shows, the $h/2e^2$ plateau appears only when $|D_R| > \sim 0.15$ V/nm. This is consistent with our prior findings on the $D$-field dependence of the LLs in bilayer graphene[19].

Resistance plateaus in $R_{34}$ appear in several other combinations of $\nu_L$ and $\nu_R$ and are consistent with the above selective complete backscattering interpretation. Figure 3(b) shows another example at $(\nu_L, \nu_R) = (-8, -4)$. A few more scenarios are given in the SM. Although the quantization in Fig. 3(b) is not as well developed as that shown in Fig. 3(a), $R_{34}$ exhibits clear plateaus or shoulders close to $h/12e^2$, $h/4e^2$, and $3h/4e^2$ expected for $\alpha = 1/4$, 1/2 and 3/4 respectively. These plateaus suggest the sequential complete backscattering of 1, 2 and 3 modes before all edge states are backscattered. As $V_{Si}$ is swept from -60 to +50 V, the four orders of magnitude change of $R_{34}$ corresponds to a change of $\alpha$ from 0 to roughly 0.9. This large tunability of $\alpha$ attests to the effective control of $V_{Si}$ on the junction potential in our devices.

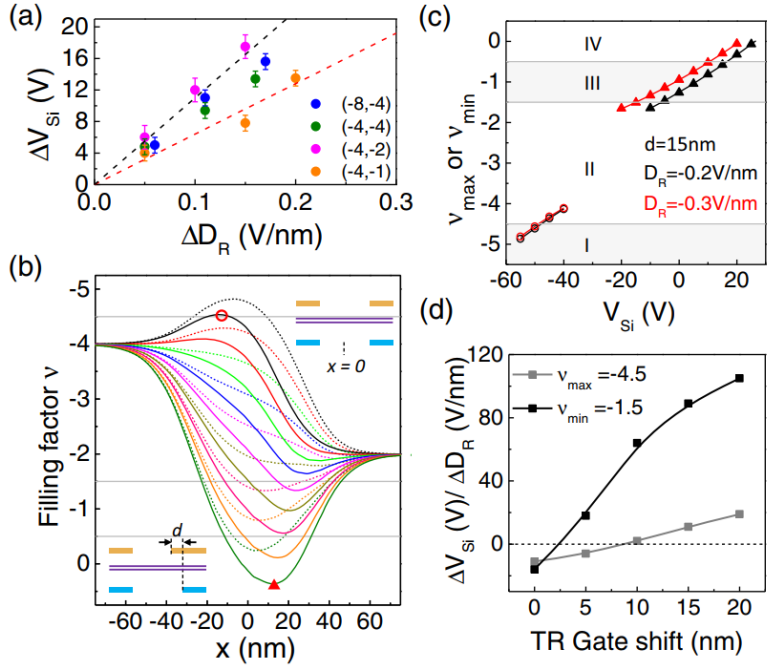

FIG. 4. (a) Measured $\Delta V_{Si}$ vs $\Delta D_R$ in different cases of $(\nu_L, \nu_R)$ as labeled in the plot. The slope of the red dashed line correspond to the value of $\Delta V_{Si}/\Delta D_R$ at $d=10$ nm on the black curve in (d). The slope of the black dashed line corresponds to $d=20$ nm. (b) Simulated filling factor profile $\nu(x)$ across the junction at selected values of $V_{Si}$ from -50 V (black) to 30 V (olive) in 10 V steps. $(\nu_L, \nu_R) = (-4, -2)$, $D_L = -0.2$ V/nm and $D_R = -0.3$ V/nm. The junction center $x = 0$ is marked in the upper inset. The dashed curves correspond to perfect gate alignment as illustrated in the upper inset. The solid curves correspond to a TR gate shift of 15nm into the junction as illustrated in the lower inset. Other dimension parameters used that of device 47. (c) $\nu_{max}$ vs $V_{Si}$ (open circles) and $\nu_{min}$ vs $V_{Si}$ (solid triangles) obtained from simulating the gate arrangement shown in the lower inset of (b) with $d = 15$nm and $D_R = -0.2$ (black symbols) and -0.3 V/nm (red symbols). $D_L = -0.2$ V/nm is fixed for both scenarios. The onset of regime II-IV corresponds to $V_{Si} = -47.7$, -6.2, and 16.6 V respectively for $D_R = -0.2$ V/nm and $V_{Si} = -48.7$, -15.1, and 10.5 V respectively for $D_R = -0.3$ V/nm. (d) Simulated $\Delta V_{Si}/\Delta D_R$ vs the TR gate shift distance $d$ for $\nu_{max} = -4.5$ and $\nu_{min} = -1.5$ as labeled in the plot.

It is interesting to note that in both Figs. 3(a) and (b), the onsets of regimes III and IV exhibit a systematic shift towards negative $V_{Si}$'s as $D_R$ becomes more negative, i.e. a positive $\Delta V_{Si}/\Delta D_R$. In Fig. 4(a), we plot the magnitude of $\Delta V_{Si}/\Delta D_R$ at several ($\nu_L$, $\nu_R$) scenarios as labeled in the plot. Meanwhile the onset of regime II appears insensitive to $D_R$. These behaviors cannot be explained by the change of the bulk LLs with $D_R$. Instead, we look to practical considerations such as the impact of misalignment between the top and bottom split gates. Using a finite element simulation tool (COMSOL, multiphysics package), we simulated the gating effect of all five gates in device 47. The results are summarized in Fig. 4 while the methods and more details are given in the SM. The carrier density profile $n(x)$ across the split junction is computed and converted to a filling factor profile $\nu(x)$. Figure 4(b) plots $\nu(x)$ at selected $V_{Si}$'s from -50 V to 30 V for two slightly different structures. The dashed lines correspond to the perfectly aligned gates shown in the upper inset of the graph. The solid lines correspond to the scenario shown in the lower inset, where the TR gate shifts into the junction by $d = 15$ nm. The overall shape of $\nu(x)$ and its evolution with $V_{Si}$ are what's expected intuitively. The $V_{Si} = 10$ V curve, for example, resembles the diagram shown in the inset of Fig. 3(a). The comparison of the $d = 0$ and 15 nm cases shows that the shift of the TR gate has a large effect on the minima of $\nu(x)$, $\nu_{min}$, which shift towards the right side but much smaller effect on the maxima of $\nu(x)$, $\nu_{max}$. We shall see that it correctly captures the behavior of $\Delta V_{Si} / \Delta D_R$ in different regimes.

To connect with experiment, in Fig. 4(c), we plot $\nu_{max}$ (open circles) and $\nu_{min}$ (solid triangles) obtained from the $d = 15$ nm curves in Fig. 4(b). Two sets of $V_{TR}$'s and $V_{BR}$'s corresponding to $D_R = -0.2$ and $-0.3$ V/nm respectively are used in the simulations and the results are plotted in black and red symbols respectively in Fig. 4(c). Using the diagram shown in the inset of Fig. 3(a), we associate the onsets of regimes III and IV with $\nu_{min} = -1.5$ and $-0.5$ respectively and similarly associate the onset of regime II with $\nu_{max} = -4.5$. This allows us to identify and label the four transmission regimes in Fig. 4(c). Their onset voltages in $V_{Si}$ agree with measurements in Fig. 3(a) very well for both $D_R$ values simulated. Indeed, a positive $\Delta V_{Si} / \Delta D_R$ is observed for the onsets of regimes III and IV while the onset of regime II remains nearly stationary. In the SM, we analyze the contribution of individual gates to provide a physical picture for the results of the simulations.

Simulations and analysis similar to that shown in Figs. 4(b) and (c) are carried out for $d = 0, 5, 10, 15$ and 20 nm and the $\Delta V_{Si} / \Delta D_R$ obtained at $\nu_{max} = -4.5$ and $\nu_{min} = -1.5$ are plotted in Fig. 4(d). Here $\Delta V_{Si} / \Delta D_R$ is calculated by linearly interpolating results at $D_R = -0.2$ and $-0.3$ V/nm. $\Delta V_{Si} / \Delta D_R$ is negligible for the onset of regime II at any $d$ values while $\Delta V_{Si} / \Delta D_R$ increases with increasing $d$ for the onset of regime III, as expected. Using the simulated $\Delta V_{Si} / \Delta D_R$ as the slope, we plot two dashed lines in Fig. 4(a). The black dashed line corresponds to $d = 20$ nm while the red dashed line corresponds to $d = 10$ nm. The majority of our data falls within the area in between. A misalignment of this magnitude is consistent with the precisions of our fabrication methods [23]. We are puzzled by the spread of the data as all of them are acquired on device 47. We note that our simple electrostatics model does not take into account the specifics of the LL structure, the shape of the density of states and the self-screening of the bilayer graphene, which may all play a role in determining the precise values of the onsets. Overall, the success of the simulation highlights an important advantage of graphene gating structures, i.e., their relatively simple and deterministic electrostatic environment. This advantage can be used to facilitate simulation-guided design of future experiments and foster a stronger connection between theory and experiment.

To summarize, we demonstrate potential-controlled transmission of quantum Hall edge states in bilayer graphene by employing independent gate controls on relevant parameters of the system. The backscattering rate is continuously tunable over a large range and sequential complete backscattering of individual edge modes is observed and well understood in numerical simulations using experimental device parameters. Our results are encouraging first steps towards building more complex nanostructures, such as an electron interferometer, to probe the charge and statistics of quasi-particles in the QH and FQH regimes of bilayer graphene.

## Acknowledgment

We thank Fan Zhang and Allan H. MacDonald for helpful discussions. Work at Penn State is supported by the NSF through NSF-DMR-1506212. Work at NIMS is supported by the Elemental Strategy Initiative conducted by the MEXT, Japan and JSPS KAKENHI Grant Number JP15K21722. Part of this work was performed at the NHMFL, which was supported by the NSF through NSF-DMR-1157490 and the State of Florida. We thank Jan Jaroszynski of the NHMFL for experimental assistance.

## Reference

[1]     A. M. Chang, Reviews of Modern Physics **75**, 1449 (2003).
[2]     V. J. Goldman and B. Su, Science **267**, 1010 (1995).
[3]     R. de-Picciotto, M. Reznikov, M. Heiblum, V. Umansky, G. Bunin, and D. Mahalu, Nature **389**, 162 (1997).
[4]     A. Stern, Annals of Physics **323**, 204 (2008).
[5]     R. L. Willett, Reports on Progress in Physics **76**, 076501 (2013).
[6]     X. Lin, R. Du, and X. Xie, National Science Review **1**, 564 (2014).
[7]     G. Moore and N. Read, Nuclear Physics B **360**, 362 (1991).
[8]     R. L. Willett, L. N. Pfeiffer, and K. W. West, Proceedings of the National Academy of Sciences of the United States of America **106**, 8853 (2009).
[9]     I. P. Radu, J. B. Miller, C. M. Marcus, M. A. Kastner, L. N. Pfeiffer, and K. W. West, Science **320**, 899 (2008).
[10]    H. Fu, P. Wang, P. Shan, L. Xiong, L. N. Pfeiffer, K. West, M. A. Kastner, and X. Lin, Proceedings of the National Academy of Sciences **113**, 12386 (2016).
[11]    W. Pan, J. S. Xia, V. Shvarts, D. E. Adams, H. L. Stormer, D. C. Tsui, L. N. Pfeiffer, K. W. Baldwin, and K. W. West, Physical Review Letters **83**, 3530 (1999).
[12]    C. R. Dean *et al.*, Nature Nanotechnology **5**, 722 (2010).
[13]    L. Wang *et al.*, Science **342**, 614 (2013).
[14]    P. Maher, C. R. Dean, A. F. Young, T. Taniguchi, K. Watanabe, K. L. Shepard, J. Hone, and P. Kim, Nature Physics **9**, 154 (2013).
[15]    A. Kou, B. E. Feldman, A. J. Levin, B. I. Halperin, K. Watanabe, T. Taniguchi, and A. Yacoby, Science **345**, 55 (2014).
[16]    P. Maher *et al.*, Science **345**, 61 (2014).
[17]    B. M. Hunt et al., Nature Communications **8**, 948 (2017).

[18] A. V. Rozhkov, A. O. Sboychakov, A. L. Rakhmanov, and F. Nori, Physics Reports **648**, 1 (2016).
[19] J. Li, Y. Tupikov, K. Watanabe, T. Taniguchi, and J. Zhu, Physical Review Letters **120**, 047701 (2017).
[20] D. K. Ki, V. I. Fal'ko, D. A. Abanin, and A. F. Morpurgo, Nano Lett. **14**, 2135 (2014).
[21] A. A. Zibrov, C. R. Kometter, T. Taniguchi, K. Watanabe, M. P. Zaletel, and A. F. Young, Nature **549**, 360 (2017).
[22] J. I. A. Li, C. Tan, S. Chen, Y. Zeng, T. Taniguchi, K. Watanabe, J. Hone, and C. R. Dean, Science, **358**, 648 (2017).
[23] J. Li, K. Wang, K. J. McFaul, Z. Zern, Y. Ren, K. Watanabe, T. Taniguchi, Z. Qiao, and J. Zhu, Nature Nanotechnology **11**, 1060 (2016).
[24] J. Li, R.-X. Zhang, Z. Yin, J. Zhang, K. Watanabe, T. Taniguchi, C. Liu, and J. Zhu, arXiv,1708.02311v1 (2017).
[25] D. Abanin and L. Levitov, Science **317**, 641 (2007).
[26] B. Huard, J. Sulpizio, N. Stander, K. Todd, B. Yang, and D. Goldhaber-Gordon, Physical Review Letters **98**, 236803 (2007).
[27] J. R. Williams, L. Dicarlo, and C. M. Marcus, Science **317**, 638 (2007).
[28] S. Matsuo, S. Takeshita, T. Tanaka, S. Nakaharai, K. Tsukagoshi, T. Moriyama, T. Ono, and K. Kobayashi, Nature Communications **6**, 8066 (2015).
[29] B. Ozyilmaz, P. Jarillo-Herrero, D. Efetov, D. A. Abanin, L. S. Levitov, and P. Kim, Physical Review Letters **99**, 166804 (2007).
[30] S.-G. Nam, D.-K. Ki, J. W. Park, Y. Kim, J. S. Kim, and H.-J. Lee, Nanotechnology **22**, 415203 (2011).
[31] F. Amet, J. R. Williams, K. Watanabe, T. Taniguchi, and D. Goldhaber-Gordon, Physical Review Letters **112**, 196601 (2014).
[32] K. Zimmermann, A. Jordan, F. Gay, K. Watanabe, T. Taniguchi, Z. Han, V. Bouchiat, H. Sellier, and B. Sacépé, Nature Communications **8**, 14983 (2017).
[33] H. Overweg *et al.*, Nano Letters **17**, 2852 (2017).
[34] Klimov, Le, Yan, Agnihotri, Comfort, Lee, Newell, and Richter, Physical Review B **92**, 241301 (2015).
[35] R. J. Haug, A. H. Macdonald, P. Streda, and K. Von Klitzing, Physical Review Letters **61**, 2797 (1988).
[36] S. Washburn, A. B. Fowler, H. Schmid, and D. Kern, Physical Review Letters **61**, 2801 (1988).
[37] W. Kang, H. Stormer, L. Pfeiffer, K. Baldwin, and K. West, Nature **403**, 59 (2000).

# Supplementary material for

**Gate-controlled Transmission of Quantum Hall Edge States in Bilayer Graphene**

Jing Li[1], Hua Wen[1], Kenji Watanabe[2], Takashi Taniguchi[2], Jun Zhu[1,3*]

[1]Department of Physics, The Pennsylvania State University, University Park, Pennsylvania 16802, USA.

[2]National Institute for Material Science, 1-1 Namiki, Tsukuba 305-0044, Japan.

[3]Center for 2-Dimensional and Layered Materials, The Pennsylvania State University, University Park, Pennsylvania 16802, USA.

*Correspondence to: jzhu@phys.psu.edu (J. Zhu)

**Online Supplementary Material Content**

1. Device characteristics
2. Landau levels of bilayer graphene in a magnetic field
3. Additional examples of gate-controlled transmission of QH edge states
4. COMSOL simulations: methods and details

## 1. Device characteristics

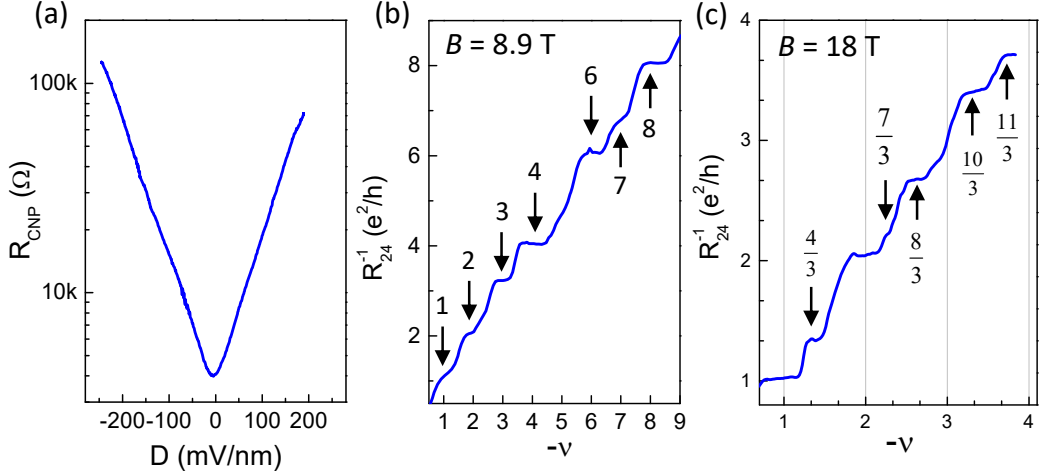

FIG. S1. (a) $R_{CNP}$ vs $D$-field in a semi-log plot. From device 47. (b) and (c) $R_{24}^{-1}$ vs filling factor $v$ measured on the right side of device 43 at $B$ = 8.9 T and 18 T respectively. $v$ is changed by sweeping $V_{TR}$. $V_{BR}$ = 1 V in (b) and -1.6 V in (c). $V_{BL} = V_{TL}$ = -3 V, $V_{Si}$ = -40 V. Arrows indicate integer and fractional fillings factors. $v$ =-3 is missing in (c) due to vanishing $D$-field at that point. $T$ = 1.6 K for (a) and (b), $T$ = 0.3 K for (c).

## 2. Landau levels of bilayer graphene in a magnetic field

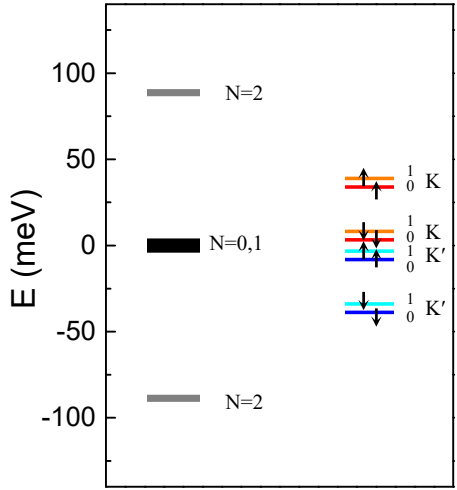

FIG. S2. A Landau level diagram of bilayer graphene. The black and gray levels show the single-particle spectrum $E = \pm\hbar\omega_c \sqrt{N(N-1)}$, where $\omega_c = eB/m^*$ is the cyclotron frequency of bilayer graphene. $m^*$ = 0.033 $m_e$ is the effective mass of bilayer graphene near $n$ = 0. At high $B$ and $D$ fields, the spin, valley and orbital degeneracies are fully lifted. On the right, we show the calculated levels for $B$ = 18 T and $D$ = 0.2 V/nm using the empirical formulae we obtained in Ref. [19] of the main text.

Although the LLs are spin, valley and orbital polarized, the voltage probes used in our measurements allow the mixing of different indices. As a result, the Landauer-Büttiker analysis yields results equivalent to a model that is index-free. In another word, our measurements are only sensitive to the number of edge states backscattered but not to their characters. This is different from the experiments in Refs. [31] and [32] in the main text, where spin-specific equilibration manifests.

## 3. Additional examples of gate-controlled transmission of QH edge states

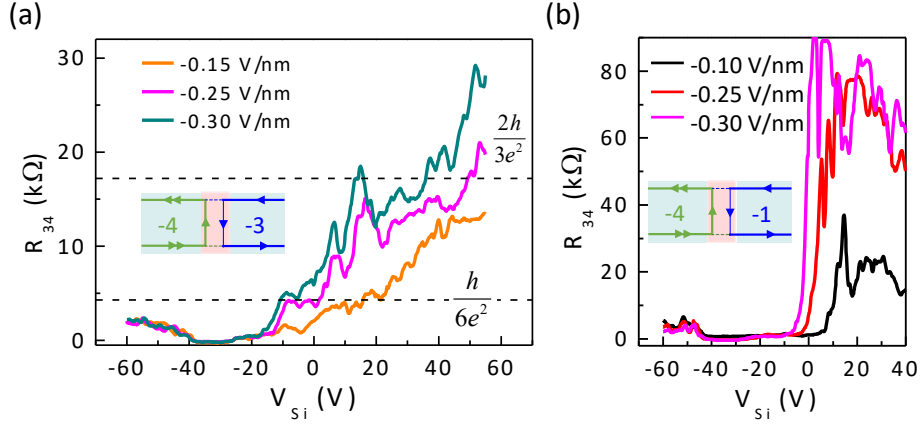

FIG. S3. (a) $R_{34}$ vs $V_{Si}$. $D_L$ = -0.2 V/nm for all traces, $D_R$ is as labeled in the plot. $(v_L, v_R)$ = (-4, -3) as the inset illustrates. Positive $D$ corresponds to an electric field pointing up. The lower dashed line indicates resistance value of $h/6e^2$ = 4.3 kΩ, which corresponds to $\alpha$ = 1/3, i. e. one of the three edge modes of the right QH states is completely backscattered while the other two fully transmit through. A shoulder at this resistance value is visible in all three traces. $\alpha$ = 2/3 corresponds to $R_{34}$= $2h/3e^2$ (the upper dashed line). The signature is less clear. (b) $R_{34}$ vs $V_{Si}$ in the case of $(v_L, v_R)$ = (-4, -1) as the inset illustrates. $D_L$ = -0.2 V/nm, $D_R$ is as labeled the plot. No intermediate plateau is observed in this case since $v_R$ only has one edge mode. In all traces, a perfect transmission regime, where $R_{34}$ is close to zero is clearly visible. Its onset in $V_{Si}$ is insensitive to $D_R$ whereas the onset of regime III (plateau) or IV (sharp rise) shifts to more negative $V_{Si}$ as $D_R$ becomes more negative. These observations are consistent with what's shown in Fig. 3 for $(v_L, v_R)$ = (-4, -2) and (-8, -4). From device 47.

## 4. COMSOL simulations: methods and details

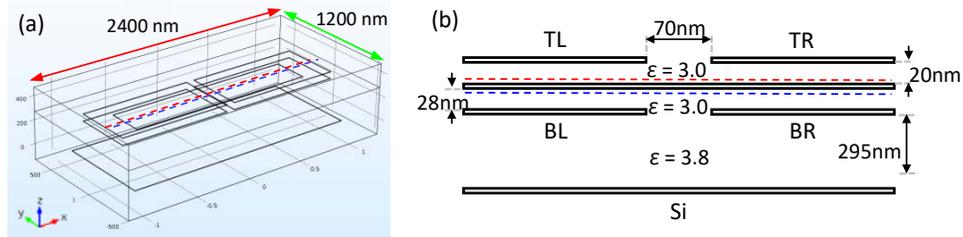

Fig. S4. (a) A schematic of the device structure constructed in COMSOL. (b) Side view of the five gates and the bilayer graphene sheet with dimensions marked in the figure (not drawn to scale). Red and blue dashed lines indicate where $D_z(x)$ is taken.

We performed detailed finite element simulations on the effect of the five gates in device 47 using the COMSOL Multiphysics software with an electrostatics package. The simulated structure is shown in Fig. S4, using parameters of device 47. The dielectric constant $\varepsilon$ of the hexagonal-Boron Nitride (h-BN) and the $SiO_2$ is set to 3 and 3.8 respectively. All five gates and the bilayer graphene sheet are represented by gold slabs of 5nm in thickness. We compute the carrier density of the top/bottom layer $n_{t/b}(x)$ along the midline of the device using the $z$ component of the displacement field $D_z(x)$ at a distance 0.1 nm above or below the bilayer graphene sheet. The net carrier density $n(x) = n_t(x) + n_b(x)$ and the filling factor profile $\nu(x) = n(x)/4.3 \times 10^{11}$ cm$^{-2}$ at $B = 18$ T. The $x = 0$ point is defined as the middle point of the bottom split gates. In the left and right bulk QH regions, the simulated gating efficiencies of the four split gates agree with measurements. Near the junction, all five gates play a role and Fig. S5 plots their gating efficiencies as a function of $x$. Solid curves correspond to perfect gate alignment. The doped Si backgate has a maximum efficiency of $g_{Si} = 2.4 \times 10^{10}$ cm$^{-2}$V$^{-1}$ at $x = 0$, which is roughly 1/3 of its planar capacitance value. $g_{Si}(x)$ decays away from $x = 0$ due to screening of the other gates. At $x = 0$, all four split gates have roughly the same gating efficiency, despite the different h-BN thicknesses (20 nm for top gates and 28 nm for the bottom gates). Thus, as long as the maximum/minimum of $\nu(x)$ remains near $x = 0$, a change of $D_R$ should not cause any significant shift in the measured $R_{34}$ ($V_{Si}$). In Fig. S6(a), we plot the simulated filling factor profile $\nu(x)$ corresponding to perfectly aligned gates for two different $D_R$ values. Indeed, $\nu_{max}$ and $\nu_{min}$ remain close to $x = 0$ and insensitive to $D_R$. This would imply the onset $V_{Si}$ of all transmission regimes discussed in the text should also be insensitive to $D_R$. On the contrary, in our measurements, only the onset of regime II remains stationary while the onsets of regimes III and IV consistently exhibit a positive $\Delta V_{Si} / \Delta D_R$ of ~ 80 V/ (V/nm) (Fig. 4(a) of the text). This discrepancy can be reconciled by allowing the TR gate to shift into the junction, as the diagram above Fig. S5 illustrates. The gating efficiency of the TR gate $g_{TR}$ corresponding to $d = 15$ nm is plotted in Fig. S5 as a red dashed line. The shift gives the TR gate stronger influence over the junction. In Fig. S6(b) and (c), we plot the simulated filling factor profile $\nu(x)$ and the $x$ coordinate of $\nu_{max}$ and $\nu_{min}$ for two different $D_R$ values in the case of $d = 15$ nm. As $D_R$ becomes more negative, $\nu_{min}$ shifts to more positive $x$ values, where $g_{TR}$ is further enhanced and $g_{Si}$ further weakened (See Fig. S3). This means the excess electron doping created by the TR gate now requires more negative $V_{Si}$ to compensate, in agreement with the trend observed in experiment. In Figs. 4(a) and (d) of the main text, we quantitatively examine $\Delta V_{Si} / \Delta D_R$ for different values of $d$ and compare to experiment. $d = 10 - 20$ nm is found to capture measurements very well.

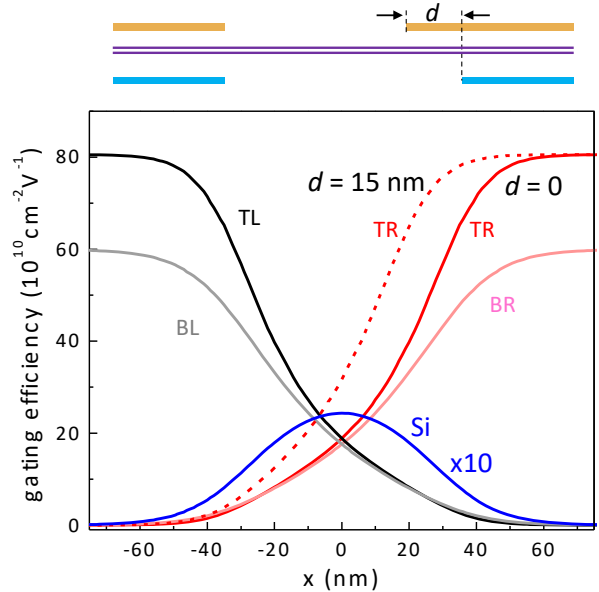

FIG. S5. Gating efficiency of individual gates as a function of $x$ as labeled in the plot. The configuration of the gates is shown above the plot with the bottom split gates starting at $x = \pm 35$ nm.

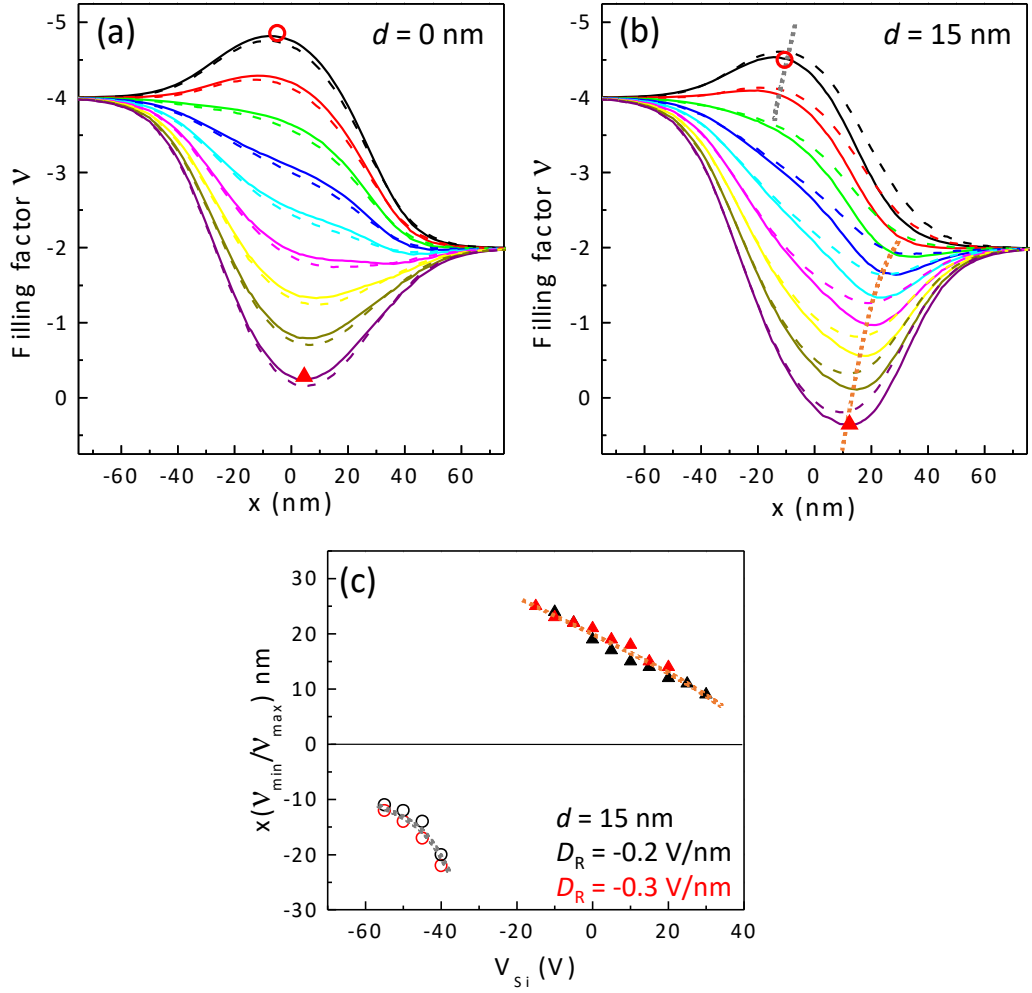

FIG. S6. (a) Simulated filling factor profile $v(x)$ across the junction at selected values of $V_{Si}$ from -50 V (black) to 30 V (purple) in 10 V steps for perfectly aligned gates. $(v_L, v_R) = (-4, -2)$ and $D_L = -0.2$ V/nm. $D_R = -0.2$ V/nm for dashed lines and -0.3 V/nm for solid lines. (b) The same as (a) but with a TR gate shift of $d = 15$ nm. (c) The $x$ coordinate of the maximum (open circles) and minimum (solid triangles) obtained from curves in (b). Grey and orange dotted lines are guides to the eye for $v_{max}$ and $v_{min}$ respectively. $v_{max}$ shifts to negative $x$ while $v_{min}$ shifts to positive $x$. The magnitude of the shift increases with increasing $|D_R|$.